\documentclass[draftcls,onecolumn,11pt]{IEEEtran}
\usepackage{cite}

\usepackage[dvips]{graphicx}

\usepackage[cmex10]{amsmath}
\usepackage{amssymb}
\usepackage{bm}
\usepackage{bbm}

\usepackage{array}
\newcommand{\va}{`}
\newcommand{\vva}{``}
\newcommand{\mc}{\mathcal}
\newcommand{\beq}{\begin{equation}}
\newcommand{\eeq}{\end{equation}}

\newcommand{\ra}{\rightarrow}

\newcommand{\qed}{\nobreak \ifvmode \relax \else
      \ifdim\lastskip<1.5em \hskip-\lastskip
      \hskip1.5em plus0em minus0.5em \fi \nobreak
      \vrule height0.75em width0.5em depth0.25em\fi}

\begin{document}
%
\title{\titolo}
%
%
%

\title{Cooperative Spectrum Sensing based on the Limiting Eigenvalue Ratio Distribution in Wishart Matrices}
\author{Federico Penna,~\IEEEmembership{Student Member,~IEEE,}
        ~Roberto Garello,~\IEEEmembership{Senior~Member,~IEEE,}\\
         and Maurizio A. Spirito,~\IEEEmembership{Member,~IEEE}
\thanks{F. Penna is with the Department of Electrical Engineering (DELEN), Politecnico di Torino, and with 
TRM Lab, Istituto Superiore Mario Boella (ISMB), Torino, Italy.
R. Garello is with DELEN, Politecnico di Torino.
M. A. Spirito is with TRM Lab, ISMB.
e-mail: \{federico.penna, roberto.garello\}@polito.it, spirito@ismb.it}
\vspace{-5mm}
}


\maketitle

\begin{abstract}
Recent advances in random matrix theory have spurred the adoption
of eigenvalue-based detection techniques for cooperative spectrum
sensing in cognitive radio. These techniques use the ratio between
the largest and the smallest eigenvalues of the received signal
covariance matrix to infer the presence or absence of the primary
signal. The results derived so far  are based on asymptotical
assumptions, due to the difficulties in characterizing the exact
eigenvalues ratio distribution. By exploiting a recent result on
the limiting distribution of the smallest eigenvalue in complex
Wishart matrices, in this paper we derive an expression for the
limiting eigenvalue ratio distribution, which turns out to be much
more accurate than the previous approximations also in the
non-asymptotical region. This result is then applied to calculate
the decision sensing threshold as a function of a target
probability of false alarm. Numerical simulations show that the
proposed detection rule provides a substantial  improvement
compared to the other eigenvalue-based algorithms.
\vspace{-3mm}
\end{abstract}



%
\IEEEpeerreviewmaketitle

\section{Introduction}

Blind detection algorithms, relying on the received signal diversity achieved through multiple antennas,
user cooperation, or oversampling, have been recently proposed for Cognitive Radio.
Most of these methods \cite{debbah, liang_pimrc} are based on the
properties of the eigenvalues of the received signal's covariance
matrix and use results from random matrix theory (RMT).

Their main advantage, with respect to classical energy detection
(ED) or cyclostationary feature detection (CFD)  \cite{akyildiz},
is that they do not require any prior information on the primary
signal or on the noise power. Among blind algorithms, the
eigenvalue-based approach was shown to outperform ED, especially
in case of noise uncertainty \cite{liang_pimrc}.

However, the decision rules of the eigenvalue-based detection
schemes proposed so far are based on asymptotical approximations,
that make them inaccurate in many practical scenarios. Using some
recent RMT results, in this paper we first derive an analytical
expression for the limiting distribution of the ratio between the
largest and the smallest eigenvalues of the covariance matrix.
Then, based on this result, we obtain a novel decision rule that
outperforms the previously proposed eigenvalue-based detection
schemes.

The rest of the paper is organized as follows: Sec. \ref{bm}
reviews the eigenvalue-based algorithms proposed in the
literature. Sec. \ref{contrib} deals with the threshold
optimization problem and presents the contribution of this paper.
Numerical results are presented and discussed in
Sec. \ref{res}. Sec. \ref{conclusion} contains the conclusions.

\section{Mathematical Background}
\label{bm}

\subsection{System Model}
\label{model}

Denote with $K$ the number of collaborating  receivers (or
antennas) and with $N$ the number of samples collected by each
receiver during the sensing time; let $y_k(n)$ be the discrete
baseband sample at receiver $k$ ($k=1,\hdots,K$) and time instant
$n$ ($n=1,\hdots,N$). Two hypotheses exist: under $\mathcal H_0$
(no primary signal: the samples contain only noise)
$y_k(n)|\mathcal{H}_0 = v(n)$, where $v(n)$ is circularly
symmetric complex Gaussian (CSCG) noise with zero mean and variance
$\sigma^2_v$; under $\mathcal H_1$ (presence of primary signal)
$y_k(n)|\mathcal{H}_1=h_k(n) s(n)+v(n)$, where $s(n)$ is the
primary signal, with $\mathrm{E} |s(n)|^2=\sigma^2_s \neq 0$, and
$h_k(n)$ is the channel between primary source and receiver $k$ at
time $n$.

Let $\mathbf{y}(n)=\left[ \, y_1(n) \hdots y_K(n) \right]^T$ be a $K\times 1$ vector containing $K$ received samples at time $n$ and $\mathbf{Y} = \left[\, \mathbf{y}(1) \hdots \mathbf{y}(N) \right]$ a $K\times N$ matrix containing all the samples received during the sensing period.
The sample covariance matrix, $\mathbf{R}(N)=\frac{1}{N} \mathbf{Y}\mathbf{Y}^H$, converges to $\mathbf{R}=\mathrm{E}[\, \mathbf{y}\mathbf{y}^H]$ for $N\rightarrow \infty$: from the eigenvalues of $\mathbf{R}(N)$ it is possible to infer the presence or absence of primary signal.

\subsection{Previous Results}
\label{maxmin}

Let $\lambda_{max}$ and $\lambda_{min}$ be the largest and the
smallest eigenvalues of $\mathbf{R}(N)$, and $l_{max}$ and
$l_{min}$ those of the normalized covariance matrix, defined as
$\mathbf{R}'(N) = \frac{N}{\sigma_v^2}\mathbf{R}(N)$. Under
$\mathcal H_0$, $\mathbf{R}'(N)$  turns out to be a complex white
Wishart matrix and, by the Marchenko-Pastur law, the eigenvalue
support is finite \cite{bai2}. 
Under $\mc{H}_1$, the covariance matrix belongs to the class of \va spiked population models'  and its largest eigenvalue increases outside the Marchenko-Pastur support
\cite{baik}. 
This property suggests to use
$T=l_{max}/l_{min}=\lambda_{max}/{\lambda_{min}}$ as test
statistic for signal detection. Denoting as $\gamma$ the decision
threshold, the detector decides for $\mc{H}_0$ if $T<\gamma$, for
$\mc{H}_1$ otherwise. Two approaches to set $\gamma$ are proposed
in the literature.

\subsubsection{Asymptotic Approach \cite{debbah}}
\label{asymp} 

Thanks to the asymptotical properties of Wishart matrices \cite{bai2}, the smallest and the largest eigenvalues of $\mathbf{R}'(N)$ under $\mc H_0)$ converge almost surely to
\begin{eqnarray}
\label{lmin}
l_{min} \ra  a = \left( N^{1/2} - K^{1/2} \right)^2  \\
l_{max} \ra  b = \left( N^{1/2} + K^{1/2} \right)^2
\label{lmax}
\end{eqnarray}
in the limit
\beq
\label{regime}
N,K \ra \infty \; \; \mbox{with} \; \; K/N \ra \overline c
\eeq
where $\overline{c} \in (0,1)$ is a constant.
Under $\mathcal H_1$, according to the theory of spiked models, the largest
eigenvalue converges almost surely to a value $b'>b$ \cite{baik}. Based on
these results, an asymptotic detection rule was proposed in \cite{debbah} with decision threshold
\begin{equation}
\gamma_{as}=\frac{b}{a}
\end{equation}

\subsubsection{Semi-asymptotic Approach \cite{liang_pimrc}}
\label{twap} This approach
is
based on the use of the recently-found limiting distribution of
$l_{max}$ instead of its asymptotical value (\ref{lmax}). Results
from \cite{jo} state that under the same assumptions (\ref{regime})
the random variable
\begin{equation}
\label{Lmax}
L_{max} = \frac{l_{max}-b}{\nu}
\end{equation}
with
\begin{equation}
\nu=\left( N^{1/2} + K^{1/2} \right)    \left( N^{-1/2} + K^{-1/2} \right)^{1/3}
\end{equation}
converges in distribution to the Tracy-Widom law\footnote{The Tracy-Widom distribution was defined in \cite{tw} as: \\$F_{TW2}(s) = \exp \left( -\int_s^{+\infty} (x-s) q^2(x) dx \right) $, where $q(s)$ is the solution of the Painlev\'e II differential equation $q''(s) = s q(s) + 2 q^3(s)$ satisfying the condition $q(s) \sim -Ai(s)$ (the Airy function) for $s \rightarrow +\infty$. For its importance in RMT this distribution has been extensively studied and tabulated; a Matlab routine to compute is available at \cite{internet}.} of order 2.
The authors of \cite{liang_pimrc} exploit this result to link the decision threshold to the probability of false alarm, defined as
\begin{equation}
\label{pfa}
 P_{fa}=P(T>\gamma | \mc{H}_0)
\end{equation}
by using the asymptotical limit (\ref{lmin}) for the smallest eigenvalue and the Tracy-Widom cumulative distribution function (CDF) for the largest one. The threshold can be written as:
\begin{equation}
\label{gammatw}
\gamma_{sa} = \gamma_{as} \cdot\left( 1+ \frac{(\sqrt{N}+\sqrt{K})^{-2/3}}{(NK)^{1/6}} F_{TW2}^{-1}(1-P_{fa}) \right)
\end{equation}
where $F_{TW2}^{-1}(y)$ is the inverse Tracy-Widom CDF of order 2.

\section{Eigenvalue Ratio Distribution and New Detection Threshold}
\label{contrib}

The asymptotic approach (Sec. \ref{asymp}) uses limiting approximations, valid for very large
$N$ and $K$. In practical conditions,
that may be characterized by small number of observations due to
time-varying channel and$/$or detection in the shortest possible time, the asymptotic threshold turns out to be
very unbalanced with respect to the actual eigenvalue ratio distribution (see next section, Fig. \ref{cdfAll}).
In addition, this approach
does not allow to tune the threshold as a function of a target
$P_{fa}$. The semi-asymptotic approach (Sec. \ref{twap}) allows such a
control, but it is still based on the asymptotical limit
for the smallest eigenvalue and it becomes inaccurate when $N$
decreases.

Recently, Feldheim and Sodin \cite{small}
found that the smallest eigenvalue also converges to
to the Tracy-Widom distribution as $K,N\rightarrow \infty$, up to a proper
rescaling factor. Thus, the random variable:
\begin{equation}
\label{Lmin}
L_{min} = \frac{l_{min}-a}{\mu}
\end{equation}
converges in distribution to the Tracy-Widom law of order 2, with:
\begin{equation}
\mu=\left( K^{1/2} - N^{1/2} \right)   \left( K^{-1/2} -
N^{-1/2} \right)^{1/3}
\end{equation}
As a consequence of (\ref{regime}), $\mu$ is always negative in the considered range of $\overline{c}$.
Now, the test statistic $T$ may be written as:
\begin{equation}
T = \frac{l_{max}}{l_{min}} = \frac{\nu L_{max} + b}{\mu L_{min} + a}
\end{equation}

Denote with $\overline{f}_{l_{max}}(z)$ and $\overline{f}_{l_{min}}(z)$, respectively, the limiting
probability density functions (PDFs) of the numerator and the denominator of $T$ for $K, N \rightarrow \infty$. From (\ref{Lmax}) and (\ref{Lmin}), these PDFs may be expressed through a linear random variable
transformation of the second-order Tracy-Widom PDF, $f_{TW2}(x)$:
\begin{equation}
\overline{f}_{l_{max}}(z) = \frac{1}{\nu} f_{TW2}\left( \frac{z-b}{\nu} \right)
\end{equation}
and, recalling that $\mu<0$:
\begin{equation}
\overline{f}_{l_{min}}(z) = \frac{1}{|\mu|} f_{TW2}\left( \frac{a-z}{|\mu|} \right) = -\frac{1}{\mu} f_{TW2}\left( \frac{z-a}{\mu} \right)
\end{equation}
Finally, assuming $\overline{f}_{l_{max}}(z)$ and $\overline{f}_{l_{min}}(z)$ as independent (which is reasonable
for limiting distributions, with the size of $\bm R'(N)$ tending to
infinity) and applying the ratio distribution formula \cite{ratiodist},
we can write the PDF of $T$ as:
\begin{align}
\overline{f}_{T|\mc H_0}(t) & = \left[ \int_{-\infty}^{+\infty} |x| \overline{f}_{l_1,l_K}(tx,x) dx  \right] \cdot {I}_{\{t>1\}} \nonumber \\
& = \left[ \int_{0}^{+\infty} x \overline{f}_{l_1}(tx) \overline{f}_{l_K}(x) dx  \right] \cdot {I}_{\{t>1\}}
\label{pdfT}
\end{align}
where
the lower integration limit has been changed to $0$ instead of $-\infty$, since the covariance matrix is positive-semidefinite therefore all the eigenvalues are non-negative; $I_{\{\cdot\}}$ is an indicator function, with the condition $t>1$ to preserve the order of the eigenvalues ($l_1>l_K$).


Given this new result, we can now introduce a sensing algorithm
based on this limiting eigenvalue ratio distribution.  Let
$\overline{F}_T(t)$ be the cumulative density function (CDF) corresponding to
(\ref{pdfT}). 
From (\ref{pfa}), the false alarm probability is $P_{fa}=1-\overline{F}_T(\gamma)$, for large $N$ and $K$; hence, we derive the the novel decision threshold as a function of the false-alarm
probability:
\begin{equation}
\gamma_{rd} = \overline{F}_T^{-1} (1- P_{fa})
\end{equation}

In practical applications the values of $F_T^{-1}(.)$, evaluated
numerically off-line, can be stored in a look-up table and then
used by the receiver to set the proper threshold as a function of
$N$, $K$, and the target $P_{fa}$. (Note that a look-up table or a similar approach is also
needed for implementing (\ref{gammatw}), since $F^{-1}_{TW2}$ does not have a
closed-form expression).

%
%

\section{Numerical Results}
\label{res}

\begin{figure}[bt]
    \centering
        \includegraphics[width=0.47\textwidth]{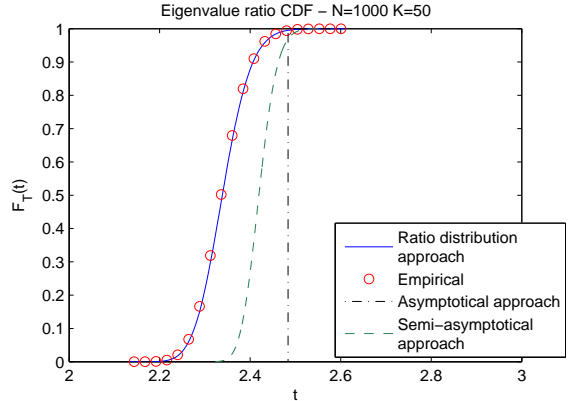}
        		\vspace{-2mm}
    \caption{Eigenvalue ratio CDF obtained using the novel ratio-based approach vs. empirical CDF, asymptotical approach, and semi-asymptotic approach. $N=1000$, $K=50$.}
    \label{cdfAll}
    		\vspace{-2mm}
\end{figure}

Fig. \ref{cdfAll} represents the eigenvalue ratio CDF resulting from the novel analytical approach and compares it to
the empirical distribution, computed by Monte-Carlo simulation, and to those obtained from the two approaches of Sec. \ref{maxmin}. The number of samples was set to $N=1000$ and the number of cooperating receivers to $K=50$. 
The novel analytical CDF matches with the empirical data, whereas
the asymptotic one (which is simply a step function) and the
semi-asymptotic one are very unbalanced because the considered
parameters ($N=1000$ samples and $K=50$ receivers), although large, are still far from the asymptotical region. From the
detector's point of view, this means that neither the asymptotic
nor the semi-asymptotic approach allow to set the decision
threshold correctly according to the target $P_{fa}$. 

\begin{figure}[bt]
        \hspace{-5mm}
    \centering
        \includegraphics[width=0.47\textwidth]{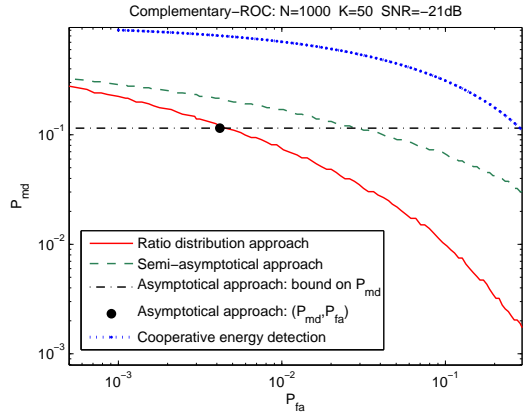}
        		\vspace{-2mm}
    \caption{Complementary ROC: the novel ratio-based approach vs. asymptotical approach, semi-asymptotic approach, and energy detection. $N=1000$, $K=50$, $SNR=-21dB$.}
    \label{roc}
    		\vspace{-2mm}
\end{figure}

Fig. \ref{roc} provides a performance comparison of the considered eigenvalue-based
detectors, plus the traditional energy detector using a cooperative equal gain combining scheme \cite{soft}. 
This type of graph, commonly used for signal
detection and called Complementary-ROC (Receiver Operating
Characteristics), represents the achievable probability of missed detection
$P_{md}=P(T<\gamma|\mc H_1)$ vs. the target $P_{fa}$.
The simulation parameters are again $N=1000$ and $K=50$;
the average signal-to-noise ratio under $\mc H_1$, defined as 
SNR $=\frac{\left\| \mathbf{h} \right\|^2 \sigma^2_s} { K \sigma^2_v} $
with $\left\| \mathbf{h} \right\|^2 = \sum_{k=1}^K \left| h_k\right|^2$, is equal to $-21$ dB.
Such low values of SNR are typically used to evaluate detectors in critical conditions (e.g., in the case of \vva hidden node'').
For energy detection, a
noise uncertainty of $0.25$ dB is assumed, whereas the
eigenvalue-based algorithms are insensitive to the noise power
uncertainty. The ROC plot shows that the novel ratio-distribution threshold provides lower probabilities of
missed detection than the other approaches for any given
probability of false alarm. 
Since the new algorithm uses a nearly-exact distribution, it allows to choose the lowest
possible threshold for a given target $P_{fa}$, i.e., to obtain
the minimum value of $P_{md}$.

For instance, given a target $P_{fa}$ of $10^{-1}$,
the novel approach provides a $P_{md}$ of $1.0 \cdot 10^{-2}$,
while the semi-asymptotic approach would give $6.5 \cdot 10^{-2}$.
The asymptotical approach, as previously mentioned, does not allow any control
of $P_{md}$ vs. $P_{fa}$ since the threshold is fixed. The pair
of $(P_{fa}, P_{md})$ it achieves is represented by a dot in the figure,
at ($4 \cdot 10^{-3}$, $1.15 \cdot 10^{-1}$); this value of $P_{md}=1.15 \cdot 10^{-1}$ is
a lower bound that cannot be improved regardless of the target $P_{fa}$, as highlighted by the straight dashed line.

\section{Conclusion}
\label{conclusion}

In this paper an expression for the limiting eigenvalue ratio distribution in Wishart matrices has been derived and it has been applied to the problem of signal detection in cognitive radio. The analytical distribution has been shown to be consistent with the empirical data and, for this reason, the novel detection rule clearly outperforms the previously proposed ones especially for realistic numbers of sensing samples and cooperative receivers.

\end{document}